\documentclass[sn-mathphys,Numbered]{sn-jnl}

\usepackage{graphicx}%
\usepackage{multirow}%
\usepackage{amsmath,amssymb,amsfonts}%
\usepackage{amsthm}%
\usepackage{mathrsfs}%
\usepackage[title]{appendix}%
\usepackage{xcolor}%
\usepackage{textcomp}%
\usepackage{manyfoot}%
\usepackage{booktabs}%
\usepackage[linesnumbered,ruled,vlined]{algorithm2e}
\DontPrintSemicolon
\usepackage{algpseudocode}%
\usepackage{listings}%
\usepackage{comment}%
\usepackage{colortbl}

\usepackage{float}
\usepackage{bbm}
\usepackage{subfig}
\renewcommand{\v}[1]{\boldsymbol{#1}}

\usepackage[linesnumbered,ruled,vlined]{algorithm2e}
\DontPrintSemicolon
\usepackage{amsfonts}
\usepackage{bbm}
\usepackage{multirow}

\begin{document}

\title[Bayesian Design with Anomalous Data]{Bayesian Design for Sampling Anomalous Spatio-Temporal Data}


\author*[1,2]{\fnm{Katie} \sur{Buchhorn}}\email{k.buchhorn@qut.edu.au}

\author[1,2]{\fnm{Kerrie} \sur{Mengersen}}\email{k.mengersen@qut.edu.au}
\author[1,2]{\fnm{Edgar} \sur{Santos-Fernandez}}\email{edgar.santosfernandez@qut.edu.au}
\author[1,2]{\fnm{James} \sur{McGree}}\email{james.mcgree@qut.edu.au}

\affil*[1]{\orgdiv{School of Mathematical Sciences}, \orgname{Queensland University of Technology}, \orgaddress{\street{George Street}, \city{Brisbane}, \postcode{4000}, \state{Queensland}, \country{Australia}}}

\affil[2]{\orgdiv{Centre for Data Science (CDS)}, \orgname{Queensland University of Technology}, \orgaddress{\street{George Street}, \city{Brisbane}, \postcode{4000}, \state{Queensland}, \country{Australia}}}


\abstract{Data collected from arrays of sensors are essential for informed decision-making in various systems. However, the presence of anomalies can compromise the accuracy and reliability of insights drawn from the collected data or information obtained via statistical analysis.
This study aims to develop a robust Bayesian optimal experimental design (BOED) framework with anomaly detection methods for high-quality data collection.
We introduce a general framework that involves anomaly generation, detection and error scoring when searching for an optimal design.
This method is demonstrated using two comprehensive simulated case studies: the first study uses a spatial dataset, and the second uses a spatio-temporal river network dataset.
As a baseline approach, we employed a commonly used prediction-based utility function based on minimising errors.
Results illustrate the trade-off between predictive accuracy and anomaly detection performance for our method under various design scenarios. 
An optimal design robust to anomalies ensures the collection and analysis of more trustworthy data, playing a crucial role in understanding the dynamics of complex systems such as the environment, therefore enabling informed decisions in monitoring, management, and response.}


\maketitle

\section{Introduction}

Optimal experimental design can be considered intelligent data collection \citep{bon2023being}, and while high-quality data underpins many functions of modern society, there remains a lack of methodologies that account for data quality within a design framework.
In-situ sensor technology has revolutionised data collection across various domains, including environmental monitoring of air, water and soil quality. 
Bayesian optimal experimental design (BOED) provides a natural mechanism for optimising and automating the data collection process, as a model-based framework aiming to maximise the amount of information gathered from such experiments.
In this paper, we develop a robust BOED framework that combines complex spatial and spatio-temporal modelling with anomaly detection methods, ensuring the data collected and analysed at design sites provide useful information (e.g. accurate predictions) as well as automated and reliable anomaly detection.\\

The principles of optimal design are evident across various domains, including economics \citep{kuhfeld1994efficient}, ecology \citep{zhang2018optimal}, social sciences \citep{myung2013tutorial, watson2017quest}, physics \citep{huan2013simulation, loredo2004bayesian}, and healthcare \citep{cheng2005bayesian}, among other quantitative fields.
Optimal experimental design serves as a means to attain maximum information with respect to an experimental objective, and is particularly important when such experiments are costly in terms of equipment and time.
Improving data quality through optimal design methodologies remains a largely unexplored area of research.
When designing microarray experiments, \citep{bolstad2004experimental} considered the quality of gene expression data alongside other image processing analyses. The research by \citep{tsou2010production} focused on optimal design in production systems. It balanced sunk costs and revenue increments due to process quality improvements, based on Taguchi cost functions.
Even in clinical trial settings, where optimal design methodologies are widely applied, there has been limited research on experimental objectives aimed at enhancing the quality and quantity of data collected from (for example) questionnaires \citep{edwards2010questionnaires}.
While there is evidence for design features that improve data completeness \citep{tourangeau2004spacing}, there is a recognised need for further research to evaluate these strategies.
In this work, we account for an often overlooked consideration: potential anomalies in the data collected from the optimal design.\\

The expanding body of research on optimal design for environmental monitoring underscores the significance of innovative statistical methodologies and optimal designs for complex ecosystems with multifaceted objectives.
In particular, \cite{fuentes2007bayesian} developed a spatial statistical methodology for designing cost-effective national air pollution monitoring networks, accounting for the nonstationary nature of atmospheric processes.
Adaptive design approaches have also been proposed for monitoring coral reefs \citep{thilan2023assessing, abeysiri2022adaptive}.
In survey design, \cite{scarpa2007benefit} employed a Bayesian approach for a multi-attribute valuation of environmental actions for landscape conservation and improvement.
Authors \citep{buxton1999optimal} presented a methodology for designing optimal solvent blends for minimising environmental impacts, balancing operational and environmental constraints.
Later, \citep{nikolopoulou2012optimal} reviewed optimal design in supply chain management for chemical processes, focusing on energy efficiency, waste management, and sustainable water management. 
These studies collectively emphasise the importance of Bayesian approaches and the necessity for balancing accuracy, cost, and environmental impact in environmental monitoring and management.
Despite the evident demand, wide-scale adoption of BOED for complex, real-world ecosystems has remained relatively limited due to significant computational challenges, particularly in adaptive scenarios or when dealing with complex models \citep{rainforth2023modern, beck2018fast}.
\\

River and stream ecosystems, which are crucial for both ecological habitats and economic activities, are increasingly threatened by climate change, pollution, and human activities. 
High-frequency data from sensors offer opportunities to understand and manage spatio-temporal dynamics of stream attributes.
Unlike many spatial applications, data collected on stream networks exhibit highly complex and multi-layered spatial dependencies, influenced by factors like climate gradients, organism mobility, and within-network physiochemical and biological processes \citep{peterson2013modelling}.
River network system modelling therefore must account for complex covariance relationships, also due to the branching network topology and unidirectional flow of water, resulting in the passive movement of materials, nutrients, and organisms downstream.
Only recently has BOED been applied to river network systems, but for a spatial model only \citep{buchhorn2022bayesian}.
In this paper we explore an extension of BOED for spatio-temporal river network models.
\\

Anomalies in sensor data can indicate critical events such as extreme weather conditions, pollution, or equipment failure. Rapid automated methods for identifying and determining the probable source of such anomalies are therefore vital in safeguarding lives and valuable assets.
In particular, distinguishing between anomalies that arise from sensor malfunctions, herein defined as technical anomalies, from extreme river events is essential.
Existing approaches for anomaly detection in river network environments include autoregressive models, e.g. autoregressive integrated moving average (ARIMA) \citep{leigh2019framework}, machine learning methods involving artificial neural networks
(ANN), random forests \citep{rodriguez2020detecting} 
and long short-term memory (LSTM) \citep{jones2022toward}.
ARIMA captures temporal dependencies only, shown to be suitable in understanding trends and cyclical patterns in water quality parameters. The machine learning techniques considered both univariate and multivariate dependencies. 
However, such methods lack the ability to provide a holistic understanding of water quality dynamics. Simultaneously accounting for spatial and temporal variation was considered in subsequent work by \citep{santosunsupervised}, showcasing the effectiveness of spatially aware models (posterior predictive distributions, finite mixtures, and Hidden Markov Models) in capturing extreme river events occurring across multiple locations. 
Also, \citep{buchhorn2023graph} use a graph neural network model to capture inter-sensor relationships as a learned graph, and employs graph attention-based forecasting to predict future sensor behaviour. However, attributing the origin of an anomaly to the correct sensor is challenging using this method.
A recent feature-based method called \emph{oddstream}, suitable for nonstationary streaming time-series data, is proposed by \citep{talagala2020anomaly}. Oddstream uses kernel density estimates on a 2-dimensional projection with extreme value theory to calculate a threshold boundary to detect outliers in time series data. \\

Bayesian design principles with optimised anomaly detection represent a novel and promising avenue for advancing the real-world application in experimental design.
In this paper, we consider the context of in-situ, online quality control systems commonly used in system monitoring. 
In complex systems like river networks, strategically placed sensors can better identify anomalous data, discrepancies can be promptly detected and addressed, ensuring the reliability of the collected data.

\subsection{Bayesian design}

BOED is a statistical approach that integrates Bayesian inference into the process of optimal experimental design to quantify (and make decisions under) uncertainty in a coherent and systematic manner.
In a broad sense, a design $\v d$ refers to a systematic arrangement and selection of variables or factors to explore (and optimise) across all possible configurations or values, denoted by the design space, $\mathcal{D}$.
The Bayesian design process begins with the formulation of a statistical model that represents the underlying system or phenomenon being studied. This model typically includes a \textit{prior distribution}, $p(\boldsymbol{\theta})$, which can be informed by previous studies, expert opinion, or theoretical considerations, and a \textit{likelihood function}, $p(\v y | \boldsymbol{\theta}, \v{d})$, which describes how the observed data are generated given the parameters.
A \textit{utility function}, $u(\cdot)$, is used to quantify the value of different experimental outcomes from a given design.
The utility function can take various forms, depending on the specific goals of the experiment, such as maximizing information gain or minimizing cost or error. In the first case, the utility function may be defined in terms of the reduction in uncertainty about the parameters \citep[e.g., decrease in the posterior variance][]{lindley1956measure, stone1959application}.
In the second case, the utility may be a function of the financial cost, time, or other resources associated with the experiment, or it may be related to the accuracy of experimental results.
The overall objective in BOED is to choose the design, $\v d$, from the design space, $\mathcal{D}$, that maximises the expected utility, which is computed as the weighted average of the utility over all possible outcomes $(\Theta, \mathcal{Y})$. The Bayesian optimal design $\v d_*$ can therefore be expressed as,
\begin{align}
\v d_* &= \mbox{arg max}_{\v d \in \mathcal{D}} \mathbb{E}[u(\v d, \boldsymbol{\theta}, \v y)] \nonumber \\
&= \mbox{arg max}_{\v d \in \mathcal{D}} \int_{\v y \in \mathcal{Y}} \int_{\boldsymbol{\theta} \in \Theta} u(\v d, \boldsymbol{\theta}, \v y)\ p(\boldsymbol{\theta}, \v y| \v d) \ d\boldsymbol{\theta}\ d\v y \label{eq:optimal_d}. 
\end{align}

It is often challenging to find Bayesian optimal designs for realistic problems, as the expected utility is typically intractable and the design space may be high-dimensional.
In the next section, we introduce techniques to handle such challenges as well as the statistical models considered throughout this paper.


\section{Background}

This section begins with a description of the motivating case study, namely optimising the design specifying the location of in-situ sensors across a river network. Details of spatio-temporal models used for river networks are given, followed by an overview of the coordinate exchange algorithm used for optimisation throughout this paper. 

\subsection{Bayesian river network model}

Generally, spatio-temporal linear regression models are formulated as:
\begin{equation}\label{eq:linear}
\v y = \mathbf{X} \boldsymbol \beta + \v v + \boldsymbol \epsilon,
\end{equation}
where $\v y$ is a stacked response vector of size $n \times 1$, with $n = S \times T$ (for $S$ spatial locations and $T$ time points), $\mathbf{X}$ is a $n \times p$ design matrix of $p$ covariates, $\boldsymbol \beta$ is a $p \times 1$ vector or regression coefficients, and $\v v$ is a stacked vector of spatio-temporal autocorrelated random effects, and $\boldsymbol \epsilon$ is the unstructured error term with, $\text{var}(\boldsymbol \epsilon) = \sigma_0^2 \bf{I}$, where $\bf{I}$ is the identity matrix and $\sigma_0^2$ is called the nugget effect.

\subsubsection{Spatial covariance for river networks}
In geostatistics, it is standard to adopt covariance models based on the Euclidean distance between two spatial locations $x_i$ and $x_j$, such as the exponential (Equation \ref{eq:euclidean_covariance}), Gaussian (Equation \ref{eq:gaussian_covariance}), and spherical functions (Equation \ref{eq:spherical_covariance}):

\begin{align}
&\text{Exponential model:} \quad  C_{\text{Euc}}(h_e; \sigma^2_e, \rho_e) = \sigma^2_e \exp\left(-\frac{3h_e}{\rho_e}\right), \label{eq:euclidean_covariance} \\
&\text{Gaussian model:} \quad  C_{\text{Euc}}(h_e; \sigma^2_e, \rho_e) = \sigma^2_e \exp\left(-3\left(\frac{h_e}{\rho_e}\right)^2\right), \label{eq:gaussian_covariance} \\
&\text{Spherical model:} \quad  C_{\text{Euc}}(h_e; \sigma^2_e, \rho_e) = \sigma^2_e \left(1 - \left(\frac{3d}{2\rho_e}\right) + \left(\frac{d^3}{2\rho_e^3}\right)\right)\mathbbm{1}\left(\frac{h_e}{\rho_e} \le 1\right) \label{eq:spherical_covariance}.
\end{align}

\noindent Here, $h_{e}$ represents the Euclidean distance between locations, $\sigma^2_e$ is the partial sill and $\rho_e$ is the spatial range parameter.
Consider two locations, $s_i$ and $s_j$, on a river network, $\mathcal{S}$.
These models may not effectively describe spatial correlation between $s_i$ and $s_j$, considering the unique branching topology, connectivity, and water flow volume of the river network within which they exist.
To address this, tail-up (Equation \ref{eq:tailup_covariance}) and tail-down (Equation \ref{eq:taildown_covariance}) models have been proposed \citep{ver2010moving}.
Exponential tail-up models restrict autocorrelation to flow-connected sites only, using spatial weights based on the flow volume and branching structure. Exponential tail-down models consider both flow-connected and flow-unconnected sites, allowing spatial dependence even between unconnected locations:

\begin{align}
C_{\text{TU}}(h; \sigma^2_u, \alpha_u) &= 
\begin{cases}
\sigma^2_u \exp\left(-\frac{3h}{\alpha_u}\right) & \text{if sites are flow-connected,} \\
0 & \text{if sites are flow-unconnected,}
\end{cases}\label{eq:tailup_covariance} \\
C_{\text{TD}}(h; \sigma^2_d, \alpha_d) &= \sigma^2_d \exp\left(-\frac{3h}{\alpha_d}\right), \label{eq:taildown_covariance}
\end{align}

\noindent where $h$ is the distance between $s_i$ and $s_j$ measured \emph{along} the stream network.
For the purely spatial case, when $T=1$ in Equation \eqref{eq:linear}, the vector of random effects $\v v$ is only spatially related with $\boldsymbol{\Sigma} = \text{cov}(\v v)$.
A covariance mixture approach for capturing the unique spatial patterns across stream networks combines these models, accounting for Euclidean (Euc), tail-up (TU) and tail-down (TD) components as follows,
\begin{equation}\label{eq:covariance_mixture}
    \boldsymbol{\Sigma} = \mathbf{C}_{\text{Euc}} + \mathbf{W} \bf{C}_{\text{TU}} + \mathbf{C}_{\text{TD}},
\end{equation}

\noindent where $\mathbf{C}_{Euc}$, $\mathbf{C}_{\text{TU}}$ and $\mathbf{C}_{\text{TD}}$ are matrices derived from their corresponding covariance functions, and $\mathbf{W}$ is a symmetric squared matrix indicating the weights between sites used to proportionally allocate (i.e. split) the tail-up moving average function at river junctions. 
Denote the vector of spatial parameters by $\boldsymbol \gamma = (\sigma^2_e, \alpha_e, \sigma^2_u, \alpha_u, \sigma^2_d, \alpha_d, \sigma^2_0)$.

\subsubsection{Spatio-temporal river network model}\label{sec:st_model}

In this spatio-temporal analysis of stream networks, we consider
repeated measures at times $t = 1, ... , T$ at fixed spatial locations $s = 1, ..., S$. 
Following \cite{santos2022bayesian}, for continuous response variables we assume the following Bayesian hierarchical model, 
\begin{equation}
p(\v y_1, \v y_2, ..., \v y_T | \boldsymbol{\theta}, \v{d}) = \prod_{t=2}^{T} p(\v y_t | \v y_{t-1}, \boldsymbol \gamma, \mathbf{X}_t, \mathbf{X}_{t-1}, \boldsymbol{\beta}, \boldsymbol \Phi, \boldsymbol{\Sigma}, \v{d}) p(\v y_1 | \boldsymbol{\theta}, \v{d}),
\end{equation}
where $\v y_1$ is the process at $t = 1$, $\v{d}$ is the design, and

\begin{align}
p(\v y_t | \v y_{t-1}, \boldsymbol \gamma, \mathbf{X}_t, \mathbf{X}_{t-1}, \boldsymbol{\beta}, \boldsymbol \Phi, \boldsymbol \Sigma, \v{d}) &= \mathcal{N}(\boldsymbol \mu_t, \boldsymbol \Sigma + \sigma_0^2 \mathbf{I}) \\
\boldsymbol \mu_t &= \mathbf{X}_t \boldsymbol \beta + \boldsymbol \Phi(\v y_{t-1} - \mathbf{X}_{t-1}\boldsymbol \beta),
\end{align}

\noindent where $\mathbf{X}_t$ is the design matrix of covariates and an intercept.
Here, we let $\boldsymbol{\theta}$ collectively represent the model parameters including spatial parameters, $\boldsymbol{\gamma}$, regression parameters, $\boldsymbol{\beta}$, and autoregressive parameters, $\boldsymbol{\phi}$.
This dynamical model assumes constant spatial correlation over time intervals and relies on first-order Markovian dependence, where $\boldsymbol \Sigma = \text{cov}(\v v_t)\ \forall t = 2,...,T$ is the $S \times S$ spatial covariance matrix, e.g., formulated by the exponential function in Equation \eqref{eq:covariance_mixture}, and 
$\boldsymbol \Phi$ is the $S \times S$ transition matrix determining the amount of temporal autocorrelation, e.g., assuming the same temporal autocorrelation for all spatial locations. In this case, the diagonal elements of $\boldsymbol \Phi$ are all equal to $\phi$ and all the off-diagonal ones are set to zero, i.e.,
\[\boldsymbol \Phi = 
\begin{bmatrix}
\phi & 0 & 0 & \cdots & 0 \\
0 & \phi & 0 & \cdots & 0 \\
0 & 0 & \phi & \cdots & 0 \\
\vdots & \vdots & \vdots & \ddots & \vdots \\
0 & 0 & 0 & \cdots & \phi \\
\end{bmatrix}.
\]

An alternative approach to model spatio-temporal autocorrelation involves constructing the full space-time covariance function.
However the dynamical model described above that models the evolution of a spatial process is more computationally efficient, by operating with spatial covariance matrices instead of joint space-time matrices, since the computational bottleneck in these methods is the inversion of large covariance matrices \citep{santos2022bayesian}. In fact, the authors of \cite{santos2022bayesian} show these two models are equivalent mathematically.

\subsection{Finding the optimal design}

Finding the Bayesian optimal design in Equation \eqref{eq:optimal_d} requires evaluation of the integrals and an approach to search the design space for the maximisation problem.
The evaluation of the expected utility, $\text{U}(\v d) := \mathbb{E}[u(\v d, \boldsymbol{\theta}, \v y)]$, is analytically intractable for the linear mixed-model in Equation \eqref{eq:linear}, as with most cases.
In practice, an approximation to the expected utility, $\hat{\text{U}}(\v d)$, is typically evaluated using Monte Carlo integration, as follows:
\begin{equation}
\hat{\text{U}}(\v d) = \frac {1}{B} \sum^{B}_{b=1} u(\v d, \boldsymbol{\theta}^{(b)}, \v y^{(b)}),
\label{eq:monte}
\end{equation}

\noindent with draws from the prior $\boldsymbol{\theta}^{(b)} \sim p(\boldsymbol{\theta})$ and then the likelihood $\v y^{(b)} \sim p(\v y | \boldsymbol{\theta}^{(b)}, \v{d})$.

For complex experimental settings, the Approximate Coordinate Exchange (ACE) \citep{overstall2017acebayes} is an effective approach in searching the design space.
The ACE method utilises the approximate coordinate exchange algorithm, and simplifies the optimisation of (an approximation to) the expected utility in high-dimensional design spaces through Gaussian process (GP) regression models. 
Assuming a GP prior, a one-dimensional emulator for the $ij$-th coordinate of $\tilde{\text{U}}(\v d)$ is constructed as the mean of the posterior predictive distribution conditioned on a few evaluations of $\hat{\text{U}}(\v d)$.
This emulation is calculated for $i=1,...,n$ and $j=1,...,k$, and then the entire process is iterated for a total of $N_1$ times. See Appendix \ref{ace_app} for further details.
To mitigate the impact of a poor emulator, the proposal design $\v{d}^*_{ij}$ is only accepted as the next design with a probability $p^*$, determined by a Bayesian hypothesis test.
Here, $p^*$ is calculated as the posterior probability that the expected utility for the proposed design, $\v d^*_{ij}$ is greater than that of the current design. 
Typically, larger Monte Carlo sample sizes are used for these tests than for the construction of the emulator, enhancing the precision of the approximation, $\hat{\text{U}}_{ij}(\v d)$.
Unless otherwise stated, we use Monte Carlo sample sizes $B_1 = 1500$ and $B_2=1000$, respectively, in the following experiments.
This approach is adept at handling generalised linear and nonlinear models, which significantly broadens the capability and practicality of BOED for a variety of objectives, providing a strong basis for our methods application. 

\section{Method}

In this section, we introduce our novel approach that incorporates anomaly detection into the Bayesian optimal design framework to improve the ability to detect anomalies across a network. 
The method is introduced generally, followed by a detailed discussion of each component within the context of environmental monitoring. Finally, an example implementation is presented which is later utilised in the case studies.

\subsection{Bayesian design with anomaly detection}

The below algorithm serves as a general schema for Bayesian design, incorporating anomaly generation and detection techniques, while introducing a dual-purpose utility aiming to balance between the objectives of the design and the efficacy of anomaly detection.

\begin{algorithm}[H]
  \KwIn{Design, anomaly generator, anomaly detector, utility function.} 
  \KwOut{Vector of utilities.}
Generate data $\v y$ (and possibly $\v y_{\text{train}}$) based on design $\v d$.\;
Generate anomalies and contaminate data $\v y$ to produce $\v y_{\text{anom}}$.\;
Detect anomalies in $\v y_{\text{anom}}$ (possibly using $\v y_{\text{train}}$), and remove (or impute) predicted anomalies in $\v y_{\text{anom}}$ to produce $\v y_{\text{cleaned}}$.\;
Evaluate performance of the anomaly detector.\;
Evaluate objective.\;
Evaluate dual-purpose utility.\;
\caption{General schema for Bayesian design with anomaly detection}
\label{algo:general_schema}
\end{algorithm}

\subsubsection{General schema}

Algorithm \ref{algo:general_schema} begins by taking inputs: the design parameters, an anomaly generator, an anomaly detector, and a utility function.
Data $\v{y}$, and optionally, a training set $\v{y}_{\text{train}}$, are generated from the likelihood function based on the given design $\v{d}$.
Then, anomalies are generated to contaminate the data $\v{y}$ to create the data variant $\v{y}_{\text{anom}}$.
Anomalies are detected given $\v{y}_{\text{anom}}$, potentially utilising $\v{y}_{\text{train}}$ for training. Detected anomalies are then removed or imputed from $\v{y}_{\text{anom}}$ to produce $\v{y}_{\text{cleaned}}$.
Finally, the utility is evaluated based on the accuracy of the anomaly detector and experimental design objective.
The following sections discusses each of these components in more detail.

\subsubsection{Anomaly generation}\label{sec:anom_gen}
While data collected in environmental systems exhibit complex patterns of spatio-temporal dependence, here we assume that sensor-related anomalies occur independently of space, but may persist over time (indicating the potential need for sensor maintenance or battery failure).
Consider an indicator matrix, $\mathbf{A}$, with elements, $a_{ti} \in \{0,1\}$, where anomalies are denoted as 1.
Persistent anomalies (i.e., consecutive anomalous observations across time) are simulated by setting $a_{\text{A};ti}, \ldots, a_{\text{A};(t+l)i} = 1$, with,
\begin{align*}
a_{\text{A};ti} \sim& \text{Bernoulli}(p_{\text{A}}) && \text{anomaly frequency}\\
l \sim& \text{Poisson}(\lambda_{\text{A}}) && \text{length of anomaly}\\
\end{align*}
In the spatial case, when $T=1$, we set $l=1$.

\subsubsection{Anomaly detection}\label{sec:anom_detect}

In this section, we introduce two different anomaly detection methods, noting that other anomaly detection algorithms may also be used \citep{ahmed2016survey, nassif2021machine, pang2021deep, chandola2009anomaly}.

\noindent\emph{Spatial anomaly detection:}
The approach described in Algorithm \ref{algo:spatial_anomaly} effectively identifies anomalies in spatial data by evaluating the mean of each point's nearest neighbours and setting thresholds based on the standard deviation of the sensor from the training data, flagging points as anomalous if they fall outside these defined bounds.

\begin{algorithm}[H]
\caption{Anomaly Detection in Spatial Data}\label{algo:spatial_anomaly}
\KwIn{training data, $\v{y}_{\text{train}}$, and test data, $\v{y}_{\text{anom}}$.}
\KwOut{Indicator vector $\v{a} \in \mathbb{R}^{n}$ for $n$ spatial locations, with $a_{i} = 1$ indicating an anomaly for sensor $i$.}
\BlankLine
\For{$i \in 1:n$}{
    Find the k-nearest neighbours of ${y}_{\text{anom}, i}$ based on Euclidean distance, $\text{NN}_k({y}_{\text{anom}, i})$.\;
    Calculate $\bar{{y}}_{i}$ as the mean of the k-nearest neighbours, $\bar{{y}}_{i} = \frac{1}{k} \sum_{y \in \text{NN}_k({y}_{\text{anom}, i})} y$.\;
    Define thresholds $\tau^+ = \bar{y}_{i} + 3\sigma_{i}$ and $\tau^- = \bar{y}_{i} - 3\sigma_{i}$, where $\sigma_i$ is the standard deviation of the training set $\v{y}_{\text{train},i}$.\;
    \eIf{$\tau^- < y_{i} < \tau^+$ }{
        Set $a_{i} = 0$\;
    }{
        Set $a_{i} = 1$\;
    }

}
\end{algorithm}

\noindent\emph{Spatio-temporal anomaly detection}: Oddstream \citep{talagala2019feature} is an anomaly detection method that uses extreme value theory and feature-based time series analysis for early detection of anomalies in non-stationary streaming time series data. 
The Oddstream method in the Appendix as Algorithm \ref{alg:Oddstream1} starts with the training phase where it extracts features, normalises them, applies PCA, estimates the probability density of the subspace, and determines the anomalous threshold therein. The online anomaly detection phase then uses the threshold to identify anomalies in `windows' of the test data.\\

\noindent\emph{Performance metrics for anomaly detection}: For assessing the performance of the anomaly detection method, we establish the following metrics based on the confusion matrix $\mathbf{M}$:
\begin{align}
\text{specificity}(\mathbf{M}) &= \frac{\text{TN}}{\text{TN} + \text{FP}}, \label{eq:spec}
\end{align}
where TN (True Negatives) represent the number of correctly identified normal instances, while FP (False Positives) denote the misclassification of normal instances as anomalies. For our experimental aim, high specificity is desirable as it means the anomaly detection algorithm is effective at correctly identifying and retaining ``good" data, based on the learned relationships between data, $\v y$, collected at design, $\v d$.
If specificity is low, the algorithm is mistakenly flagging too many normal data points as anomalies, leading to the loss of valuable information.
This can be particularly problematic in scenarios where the cost of False Positives is high, such as the travel and time costs associated with checking remote in-situ sensors.

For completeness, we also consider the following performance metrics in the assessment and comparison of the optimised designs,
\begin{align*}
\text{sensitivity}(\mathbf{M}) &= \frac{\text{TP}}{\text{TP} + \text{FN}}, \text{ and} \\
\text{accuracy}(\mathbf{M}) &= \frac{\text{TN} + \text{TP}}{\text{TN} + \text{TP} + \text{FP} + \text{FN}},
\end{align*}
where TP (True Positives) represent correctly classified anomalies and FN (False Negatives) represent anomalies incorrectly classified as normal instances. There is a trade-off between sensitivity and specificity.
Increasing specificity usually leads to a decrease in sensitivity.

Ultimately, we use the Matthews correlation coefficient \citep[MCC][]{matthews1975comparison} for an overall evaluation as it provides an informative measure of classifier performance, particularly useful in cases where the classes are imbalanced.
A larger value of the MCC indicates better performance of the classifier in terms of both sensitivity and specificity, producing a score between -1 and +1, defined by,
\begin{equation}
\text{MCC}(\mathbf{M}) = \frac{\text{TP} \times \text{TN} - \text{FP} \times \text{FN}}{\sqrt{(\text{TP} + \text{FP})(\text{TP} + \text{FN})(\text{TN} + \text{FP})(\text{TN} + \text{FN})}}. \label{eq:mcc}
\end{equation}


\subsubsection{Dual-purpose utility}\label{sec:utility}

The utility function introduced herein is a compromise between the accuracy of the posterior predictive mean at $\hat{n}$ specified locations (inverse root mean squared error; irmse), and the specificity (sp) of anomaly detection at $n$ sampled data locations.
This compromise can be expressed as the product of the two components,
\begin{align}
\text{U}({\v d}) &= \mathbb{E}[\text{u}_{\text{irmse}}({\v y}, {\v d})]  \times \mathbb{E}[\text{u}_{\text{sp}}(\mathbf{A}_{\text{true}}, \hat{\mathbf{A}}_{\text{anom}})],\label{eq:u}
\end{align}
with,
\begin{align}
\text{u}_{\text{irmse}}({\v y}, {\v d}) &= \bigg( \frac{1}{\hat{n}} \sum^{\hat{n}}_{i=1}(\mathbb{E}(\hat{y}_i| \v{y}, \hat{\v d} ) - \hat{y}_i)^2 \bigg)^{-\frac{1}{2}},\label{eq:u_irmse}\\
\text{u}_{\text{sp}}(\mathbf{A}_{\text{true}}, \hat{\mathbf{A}}_{\text{anom}}) &= \text{specificity}(\mathbf{M}), \text{ for confusion matrix $\mathbf{M}$,} \nonumber
\end{align}
where the $i$-th element of the posterior predictive mean is written as $\mathbb{E}(\hat{y}_i| \v{y}, \hat{\v d} )$ and specificity is defined in Equation \eqref{eq:spec}.
In this context, a design, $\v d$, represents the specification of physical sensor locations, while design, $\hat{\v d}$, denotes the design of prediction locations. 
Utilising the inverse root mean squared error metric implies that design scenarios with various  $\hat{n}$ will result in utilities being measured on differing scales.
Note that multiple approaches exist to compromise between these objectives, alternative methods may also be considered, including the incorporation of other anomaly detection performance metrics.

\subsection{Example implementation}

An example implementation of our general approach is specified in Algorithm \ref{algo:b}.
In line 1, we generate a true anomaly indicator matrix, $\mathbf{A}_{\text{true}}$, specifying where actual anomalies occur in the data collected at design sites, $\v y$, based on the specified anomaly generation method.
Then, we simulate anomalies on the supposed data collected, $\v y_{\text{anom}}$, by adding normally distributed noise based on the specified mean and standard deviation, in this case.
Training data is also generated from the likelihood function.
The anomaly detection algorithm is then applied to predict anomalies, given the design locations, $\v d$, anomalous data, $\v y_{\text{anom}}$, and training data, $\v y_{\text{train}}$, resulting in a predicted anomaly indicator matrix, $\hat{\mathbf{A}}_{\text{anom}}$.
The success of the anomaly detection is evaluated as $\text{u}_{\text{sp}}$, defined as specificity computed from the confusion matrix.
The predicted anomalies are then imputed from the contaminated data $\v y_{\text{anom}}$ to create $\v y_{\text{cleaned}}$.
Crucially, it is $\v y_{\text{cleaned}}$ that is used to compute the posterior predictive mean, $\mathbb{E}(\hat{\v y}| \hat{\v d}, \v y_{\text{cleaned}})$ in the approximation to $\text{u}_{\text{irmse}}$ defined in Equation \eqref{eq:u_irmse}.
In the following section, we examine the errors associated with the posterior predictive mean given $\v y$, as well as $\v y_{\text{anom}}$, and $\v y_{\text{cleaned}}$.
Finally, an approximation to the combined utility defined in Equation \eqref{eq:u} is computed as the product of the expected predictive errors (given potentially anomalous data) and the expected specificity (performance of the anomaly detection).

\begin{algorithm}[H]
  \KwIn{$\v y_{\text{train}}$ drawn from the likelihood function, anomaly generation parameters corresponding to length, $\lambda_A$, frequency, $p_{\text{A}}$, mean, $\mu_{\text{A}}$, and standard deviation, $\sigma_{\text{A}}$.} 
  \KwOut{vector of utility samples.}

\BlankLine
\For{$b \in 1:B$}{
Draw $\v y^{(b)}$ from the likelihood function.\;
Randomly generate indicator matrix, $\mathbf{A}_{\text{true}}^{(b)} | \lambda_A, p_{\text{A}}$, with entries:
\[a_{ti} = 
\begin{cases}
1, & \text{if element } (t, i) \text{ represents a true anomaly}, \\
0, & \text{otherwise},
\end{cases}
\]
for $t = 1, \ldots,  T$ and $i = 1, \ldots, n$ per Section \ref{sec:anom_gen}.\;
Randomly generate proposed contaminated data, $\v{y}^{(b)}_{\text{anom}} | \mu_{\text{A}}, \sigma_{\text{A}}$, as:
\begin{equation*}
y_{\text{anom};ti} = 
\begin{cases}
y_{ti}, & \text{if } a_{ti}=0, \\
y_{ti} + z, & \text{otherwise, with } z \sim \mathcal{N}(\mu_{\text{A}}, \sigma_{\text{A}}^2)
\end{cases}
\end{equation*}
for elements $a_{ti} \in \mathbf{A}_{\text{true}}^{(b)}$ and $y_{ti} \in \v y^{(b)}$.\;
Detect anomalies via a method outlined in Section \ref{sec:anom_detect}; obtain predicted anomaly indicator matrix $\hat{\mathbf{A}}_{\text{anom}}^{(b)}$. \;
Remove anomalies to obtain $\v y^{(b)}_{\text{cleaned}}$ for which the corresponding elements of $\hat{\mathbf{A}}_{\text{anom}}^{(b)}$ equal 0.\;
}

Compute combined utility which can be approximated as,
 \[{\text{U}}(\v d) \approx \hat{\text{U}}_{\text{irmse}}({\v d}) \times \hat{\text{U}}_{\text{sp}}(\mathbf{A}_{\text{true}}, \hat{\mathbf{A}}_{\text{anom}}), \]
via Monte Carlo approximation,
$$\hat{\text{U}}_{\text{irmse}}({\v d}) = \frac{1}{B} \sum^{B}_{b=1} \text{u}_{\text{irmse}}({\v y^{(b)}_{\text{cleaned}}}, {\v d}) .$$
and with $\mathbf{A}_{\text{true}} = \odot^{B}_{b=1}\mathbf{A}^{(b)}_{\text{true}}$ and 
$\hat{\mathbf{A}}_{\text{anom}} = \odot^{B}_{b=1} \hat{\mathbf{A}}_{\text{anom}}^{(b)}$
where $\odot$ represents vertical concatenation.


  \caption{Implementation of Bayesian design with anomaly detection}
  \label{algo:b}
\end{algorithm}

\section{Case studies}
In this section, we investigate the proposed method on a spatial example and then our motivating spatio-temporal river network example, each with a corresponding anomaly detection method.

\subsection{Spatial simulation}

In this design problem, we consider $n=6$ spatial locations $\v x_i = (x_{i1}, x_{i2})^\top$ within a specified 2-dimensional area for observing responses $\v y = (y_1,...,y_n)^\top$. A Gaussian process model is fit to $\v y$ for prediction at $\hat n$ predefined unobserved locations $\hat{\v x_i} = (\hat{x}_{i1}, \hat{x}_{i2})^\top$, corresponding to responses $\hat{\v y} = (\hat{y}_1,...,\hat{y}_{\hat{n}})^\top$. The combined vector of observed and unobserved responses is denoted as $\tilde{\v y} = (\v y^\top, \hat{\v y}^\top)^\top$, where $\tilde{n} = n + \hat{n}$. The objective is to optimise the design $\v d := \{ \v x_i\}_{i=1}^n$ for effective prediction of $\hat{\v y}$, despite anomalies in $\v y$. 
The model is a zero-mean Gaussian process, described by:
\begin{equation}
p(\tilde{\v y} | \sigma^2, \rho, \tau^2) \sim \mathcal{N}\big(\v 0, \sigma^2 \tilde {\boldsymbol \Sigma} \big).
\label{eq:spatial}
\end{equation}
\noindent Here, $\tilde{\boldsymbol \Sigma} = \tilde{\mathbf{C}} + \tau^2 \mathbf I_{\tilde n}$, with $\tilde n \times \tilde n$ identity matrix $\mathbf{I}_{\tilde n}$, and $\tilde{\mathbf{C}}$ being an $\tilde n \times \tilde n$ correlation matrix partitioned as follows:
\[\tilde {\mathbf{C}} = \begin{bmatrix}
{\mathbf{C}} & {\mathbf{S}}\\
{\mathbf{S}}^\top & {\bf \hat{C}}
\end{bmatrix}.\]
\noindent In this partition, ${\mathbf{C}}$ is the correlation matrix for design locations $\v x_i$, ${\bf \hat{C}}$ for prediction locations $\hat{\v x_i}$, and ${\mathbf{S}}$ between locations $\v x_i$ and $\hat{\v x_i}$. 
The elements of $\tilde{\mathbf{C}}$ are based on $C_{\text{Euc}}(h_e; \sigma^2, \rho)$ in Equation \eqref{eq:euclidean_covariance}, with priors:
\begin{align*}
\sigma^{-2} &\sim \text{Gamma}\big(\frac{3}{2}, \frac{1}{2} \big)\\
\rho &\sim \text{Uniform}\big(1, \frac{3}{2} \big)
\end{align*}
A small, fixed noise term $\tau = 1 \times 10^{-5}$ is assumed. The posterior predictive mean is given by:
\begin{equation}
\mathbb{E}(\hat{\v y}| \hat{\v d}, \v y) = {\mathbf{S}}^\top({\mathbf{C}} + \tau^2 \mathbf I_n)^{-1} \v y.
\label{rq:post_mean}
\end{equation}

\begin{figure}
\centering
\includegraphics[width = 0.6\textwidth]{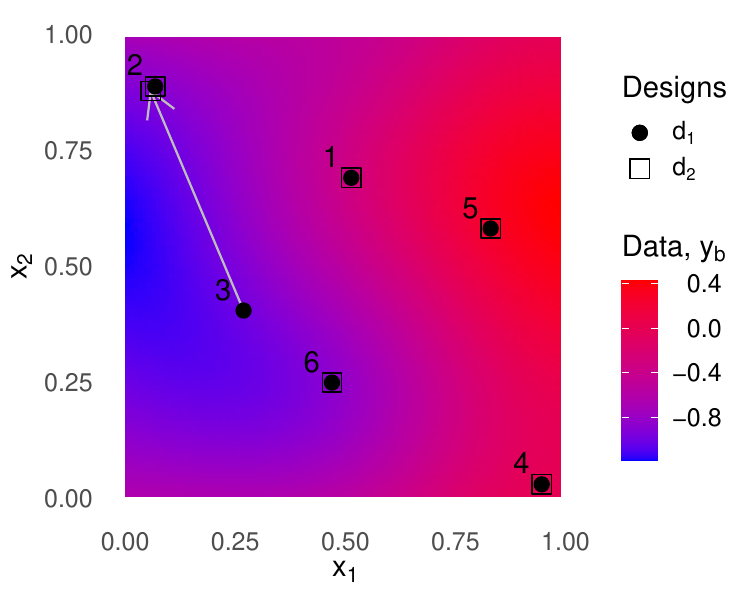}
\caption{Random design configuration, $\v d_1$, and an adjusted configuration, $\v d_2$, where sensor 3 is relocated closer to sensor 2. The background is coloured by a realisation of the Gaussian process generated by Equation \ref{eq:spatial}.
} \label{fig:spatial_locations}
\end{figure}

\begin{figure}[h]
\includegraphics[width = \textwidth]{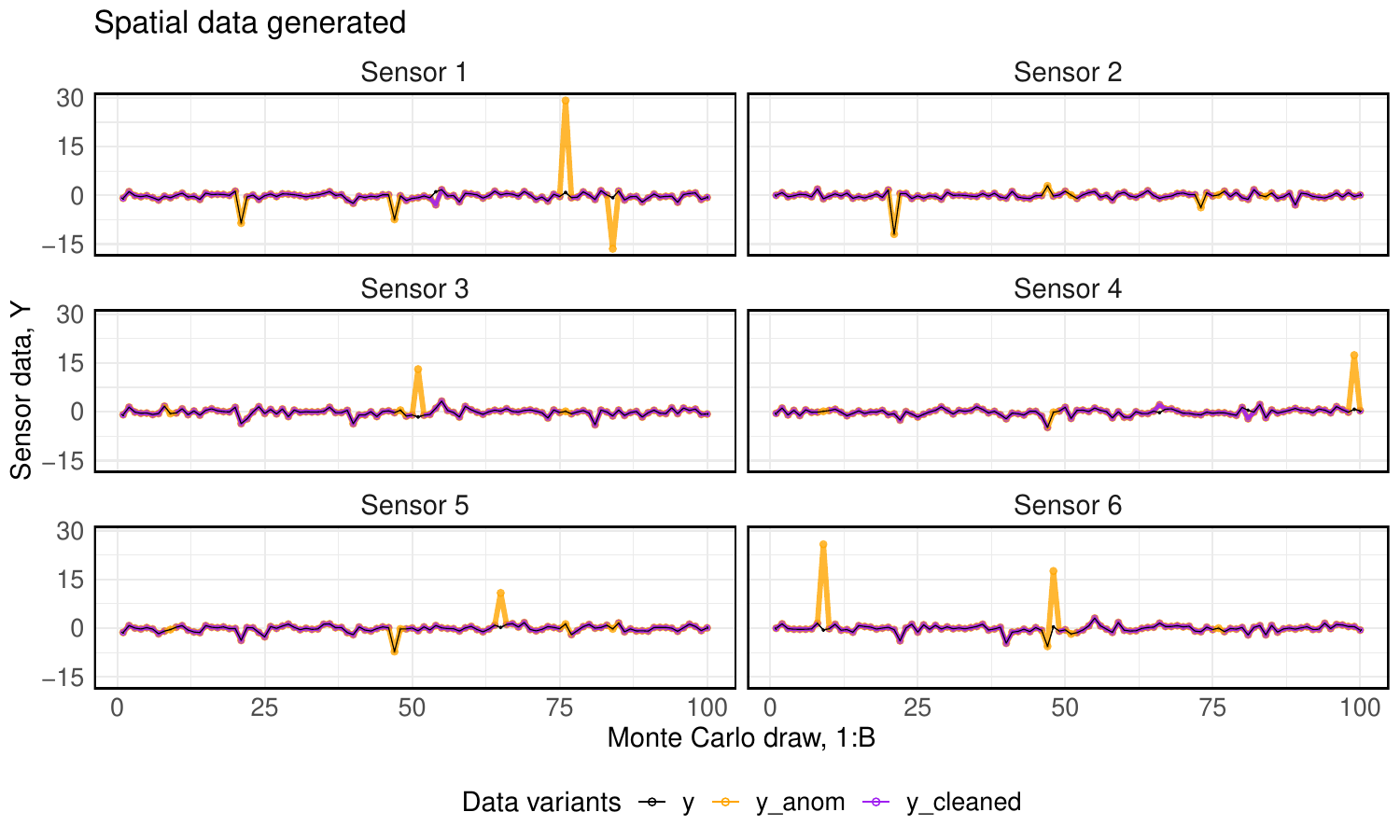}
\caption{Variation in data sets $\v y$, $\v y_{\text{anom}}$, and $\v y_{\text{cleaned}}$ generated in Algorithm \ref{algo:b} for the spatial model, and the anomaly detection method described in Algorithm \ref{algo:spatial_anomaly}. The sensor locations are specified by $\v d_1$ in Figure \ref{fig:spatial_locations}. Sensor 2 shows a false positive rate of 6\%, with an overall anomaly detection specificity of 96.4\%.} \label{fig:spatial_data}
\end{figure}

Consider a random starting configuration of the sensors, $\v d_1$, shown in Figure \ref{fig:spatial_locations}.
Anomalies are generated according to Section \ref{sec:anom_gen} with $p_{\text{A}} = 0.10$, $\mu_{\text{A}} = 5$, and $\sigma_{\text{A}}^2 = 10$.
For each design, training data with is drawn independently from the likelihood function, in this example we set $B_{\text{train}}=100$.
As per Algorithm \ref{algo:b}, the approximation to the expected utility is $\hat{\text{U}}(\v d) = 5.59$, corresponding to an anomaly detection specificity of 96.4\%.
Data generated at $\v d_1$ is shown in Figure \ref{fig:spatial_data} across the first $B=100$ draws of the prior distribution, noting that sensor 2 accounts for the largest false positive rate of 6\% among the sensors.
Adjusting the position of sensor 3 to be nearer to sensor 2 effectively reduces the occurrence of false positives in this scenario, shown in Figure \ref{fig:spatial_locations} as $\v d_2$.
The design $\v d_2$ reduces the rate of false positives in Sensor 2, corresponding to an overall anomaly detection specificity of 97.8\%  but decreases the utility approximation to $\hat{\text{U}}(\v d) = 4.98$ (due to an increase in false negatives).

\begin{figure}[h]
\includegraphics[width = \textwidth]{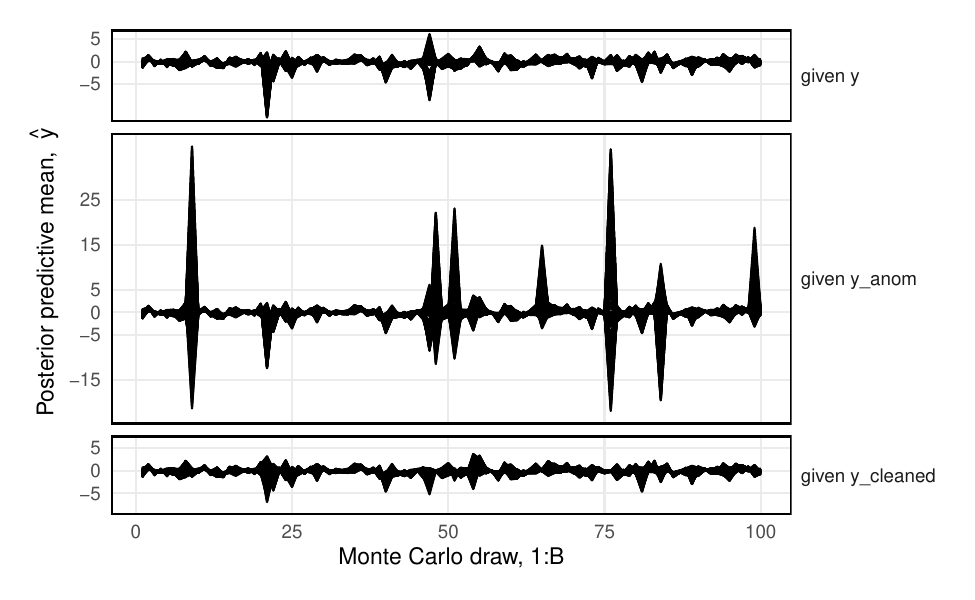}
\caption{Illustration of posterior predictive means at specified $\hat{n}$ prediction locations using data collected at $\v d_1$ under different quality conditions. The top section reflects the assumption of perfect data collection, $\mathbb{E}(\hat{\v y}| \hat{\v d}, \v y)$, the middle shows the impact of anomalies in the data, $\mathbb{E}(\hat{\v y}| \hat{\v d}, \v y_{\text{anom}})$, and the bottom demonstrates the posterior predictions after automatic anomaly removal, $\mathbb{E}(\hat{\v y}| \hat{\v d}, \v y_{\text{cleaned}})$, highlighting the influence of data quality on prediction accuracy.}\label{fig:posterior_pred}
\end{figure}

We further examine the data variations $\v y$, $\v y_{\text{anom}}$ and $\v y_{\text{cleaned}}$ generated in Algorithm \ref{algo:b}, shown in Figure \ref{fig:spatial_data}.
Traditional optimal design procedures operate under the assumption that the data collected is anomaly-free. This perfect data is then used for making predictions, as indicated by the posterior predictive mean, $\mathbb{E}(\hat{\v y}| \hat{\v d}, \v y)$. However, practical data collection experiences suggest that this assumption is often flawed.
Figure \ref{fig:posterior_pred} illustrates the posterior predictive mean for the predefined $\hat{n}$ prediction locations, based on data of varying quality collected at $\v d_1$.
The top part of the figure demonstrates the scenario under the assumption of perfect data.
The middle set of posterior predictions, $\mathbb{E}(\hat{\v y}| \hat{\v d}, \v y_{\text{anom}})$, represents a real-life situation where anomalies in the measured data visibly affect the posterior predictions.
The bottom part of the figure displays the posterior predictions, $\mathbb{E}(\hat{\v y}| \hat{\v d}, \v y_{\text{cleaned}}, \hat{\v d})$, following the automatic detection and removal of anomalies in Algorithm \ref{algo:b}, showcasing the potential improvements in data quality and prediction accuracy.

\begin{figure}[h]
\begin{tabular}{ccc}
\subfloat[ACE trace plot for utility, $\text{u}$ (proposed method).]{\includegraphics[width = 0.3\textwidth]{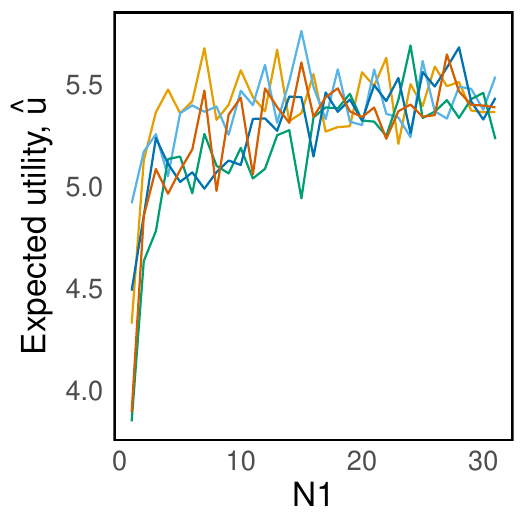}} &
\subfloat[ACE trace plot for utility, $\text{u}_{\text{irmse}}$ (prediction errors only).]{\includegraphics[width = 0.3\textwidth]{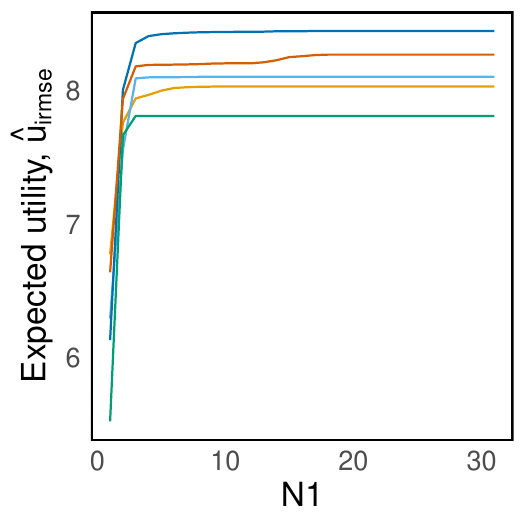}} &
\subfloat[ACE trace plot for utility, $\text{u}_{\text{sp}}$ (anomaly detection performance only).]{\includegraphics[width = 0.3\textwidth]{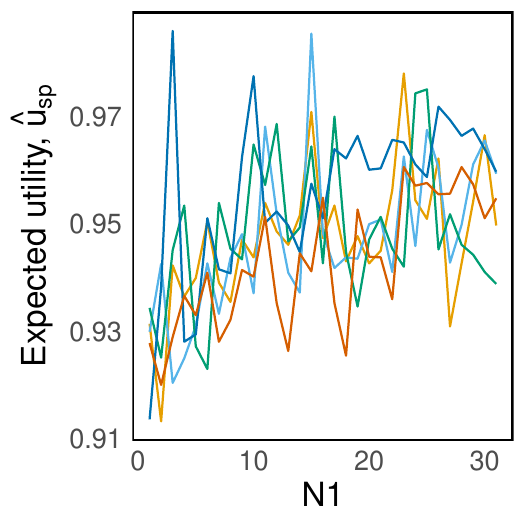}} \\
\subfloat[Prediction design, $\hat{\v d}$, and the optimal design, $\v d_*$.]{\includegraphics[width = 0.3\textwidth]{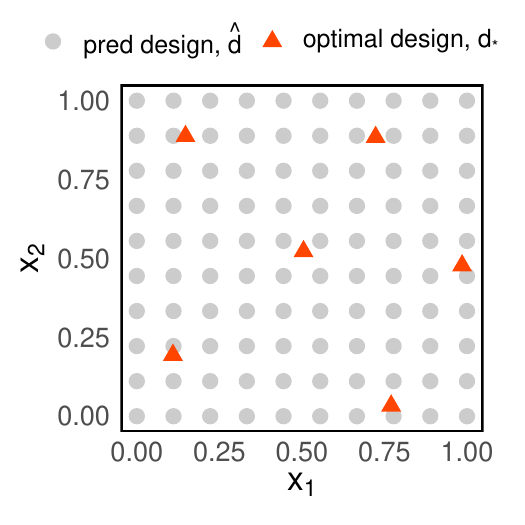}} &
\subfloat[Prediction design, $\hat{\v d}$, the optimal design, $\v d_{*\text{irmse}}$.]{\includegraphics[width = 0.3\textwidth]{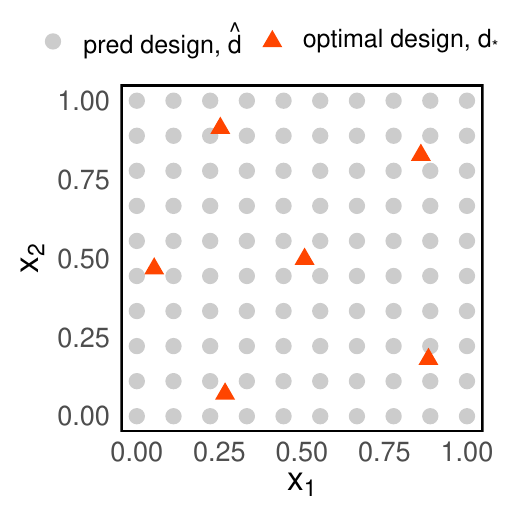}} &
\subfloat[Optimal design, $\v d_{*\text{sp}}$.]{\includegraphics[width = 0.3\textwidth]{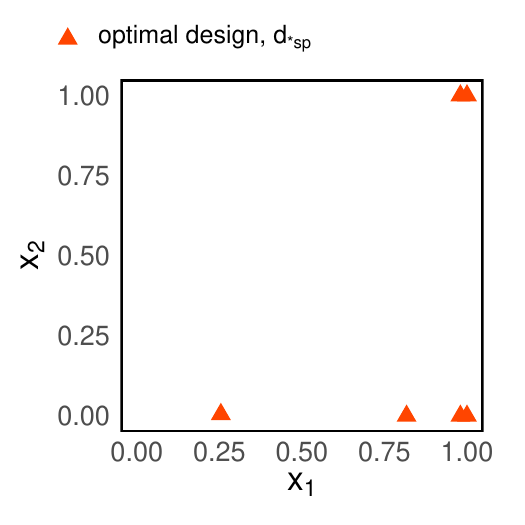}}
\end{tabular}
\caption{Top row: trace plots for the Approximate Coordinate Exchange (ACE) algorithm, optimised according to their respective utilities, defined in Equation \ref{eq:u}. Coloured lines indicate different random starts. Bottom row: spatial locations of corresponding optimal design points within the design space, and the prediction design locations.} \label{fig:trace_locations}
\end{figure}

We ran the ACE algorithm with $N_1 = 30$, utilising the proposed dual-purpose utility function from Algorithm \ref{algo:b}, to find the optimal design $\v d_*$.
The optimal design process outlined above was carried out under various scenarios of anomaly contamination, as detailed in Table \ref{tab:spatial_utility}.
For brevity, we refer to non-anomalous data as ``good" data.
This table displays the percentage of anomalies in each scenario, the corresponding percentage of anomalies removed, and the percentage of good data retained.
The findings demonstrate successful removal of most anomalies (67.7\%-80.4\%) and high retention of good data (94.0\% - 99.3\%), enhancing the overall quality of the retained data set.

The trace plots for the ACE optimisation are displayed in Figure \ref{fig:trace_locations}. 
As a baseline method, we executed the optimisation algorithm outlined above for minimising prediction error, $\text{u}_{\text{irmse}}$, resulting in the optimal design, $\v d_{*\text{irmse}}$.
For comparative purposes, we also optimised only for anomaly detection performance, $\text{u}_{\text{sp}}$, resulting in optimal design, $\v d_{*\text{sp}}$.
The locations of these optimal designs are depicted in Figure \ref{fig:trace_locations}, with anomalies generated at a proportion of 10\% (Scenario 3).
The design locations for $\v d_{*\text{irmse}}$ appear evenly spaced, consistent with other patterns for a utility based on prediction accuracy.
The observed clustering of the design locations $\v{d}_{*\text{sp}}$ is associated with the anomaly detection process, which relies on neighbouring site values to compute the expected mean.
We also observe less uniform spacing in $\v{d}_{*}$ compared to the baseline optimal design, $\v{d}_{*\text{irmse}}$.

\begin{table*}[h]
\caption{\small Definition of anomaly detection scenarios defined by the proportion of generated anomalies, $p_{\text{A}}$. The expected utility, percentage of anomalies removed, and percentage of good data retained is reported for each scenario.}\label{tab:spatial_utility}
\normalsize
\begin{tabular*}{\textwidth}{@{\extracolsep\fill}lccccc}
 \toprule
 \multirow{3}{*}{Scenario} & \multirow{3}{*}{$p_{\text{A}}$ } & \multirow{3}{*}{\% anomalies} & \multicolumn{3}{c}{$\v{d}_{*}$} \\ 
 \cmidrule{4-6}
 \rule{0pt}{3ex} 
 & & & \multicolumn{1}{c}{\centering $\hat{\text{U}}(\v d_*)$} & \multicolumn{1}{p{2cm}}{\centering \% anomalies removed} & \multicolumn{1}{p{2cm}}{\centering \% good data retained} \\ 
 \midrule
 0 & 0.000 & 0.0 & 8.03 & - & 98.7 \\
 \midrule
 1 & 0.010 & 1.0 & 7.71 & 67.7 & 99.3 \\
 \midrule
 2 & 0.050 & 5.0 & 6.91 & 80.4 & 97.1 \\
 \midrule
 3 & 0.100 & 10.0 & 5.53 & 77.7 & 94.0 \\
 \bottomrule
\end{tabular*}
\end{table*}

In Table \ref{tab:rmse_spatial}, we assess the performance of our proposed optimal design, $\v d_*$,
with the comparative optimal designs in terms of prediction accuracy, $\text{u}_{\text{irmse}}$.
In columns 1-3, we assume the application of an automated anomaly detection procedure on the data at design sites to produce $\v{y}_{\text{cleaned}}$, as outlined in Algorithm \ref{algo:b}.
The last column in the table represents a practical scenario, where sensors are placed according to the baseline optimal design for minimising prediction errors, $\v d_{*\text{irmse}}$, but the data collected at these locations are anomalous, $\v y_{\text{anom}}$.

As expected, the design optimised for prediction accuracy, $\v d_{*\text{irmse}}$, demonstrates the highest prediction accuracy.
The numbers in parenthesis indicate the standard deviation of the utility values.
The first row indicates the impact that removing False Positives has on the error, since no anomalies are generated and $\v y_{\text{anom}}$ is considered perfect data (without anomalies).
In this case, the utility decreases marginally from 8.16 to 8.08.
Scenario 1 indicates the shift away from the effectiveness of the baseline design, to the effectiveness of an automatic anomaly detection procedure, at a contamination rate of 1\%.
Interestingly, the proposed optimal design, $\v d_{*}$, demonstrates near-optimal prediction accuracy when anomalies are present. 
Note that the anomaly detection design, $\v{d}_{*\text{sp}}$, shows considerably lower prediction accuracy, even compared to the last column with no anomalies removed. 
This suggests that the predictive capability is largely affected by more clustered sites.
Particularly in the last row, a contamination rate of 10\% anomalies from data collected at the baseline design, $\v d_{*\text{irmse}}$, still results in lower error than cleaned data collected at $\v{d}_{*\text{sp}}$.

\begin{table}[h!]
\caption{\small Comparison of optimal designs $\v d_*$, $\v d_{*\text{irmse}}$ and $\v d_{*\text{sp}}$, under differing levels of anomaly contamination.
The left and middle columns calculate the inverse root mean squared errors for each optimal design, using an automatic anomaly detection process. The rightmost column illustrates a real-world scenario with sensors placed according to $\v d_{*\text{irmse}}$ encountering anomalous data, $\v y_{\text{anom}}$.} \label{tab:rmse_spatial}
\normalsize 
\begin{tabular*}{\textwidth}{@{\extracolsep\fill}lcccc}
 \toprule
 \multirow{2}{*}{Scenario} 
 & \multicolumn{1}{p{2.2cm}}{\centering $\text{u}_{\text{irmse}}(\v d_{*})$ w/ \\ $\mathbb{E}(\hat{\v y}| \hat{\v d}, \v y_{\text{cleaned}})$ }
 & \multicolumn{1}{p{2.7cm}}{\centering $\text{u}_{\text{irmse}}(\v d_{*\text{irmse}})$ w/ \\ $\mathbb{E}(\hat{\v y}| \hat{\v d}, \v y_{\text{cleaned}})$ }
  & \multicolumn{1}{p{2.4cm}}{\centering $\text{u}_{\text{irmse}}(\v d_{*\text{sp}})$ w/ \\  $\mathbb{E}(\hat{\v y}| \hat{\v d}, \v y_{\text{cleaned}})$ }
  & \multicolumn{1}{p{2.7cm}}{\centering $\text{u}_{\text{irmse}}(\v d_{*\text{irmse}})$ w/ \\  $\mathbb{E}(\hat{\v y}| \hat{\v d}, \v y_{\text{anom}})$ }\\ 
 \midrule
 0 & 8.00 (4.73) & \underline{8.08} (4.80)& 2.59 (1.66)& \textbf{8.16} (4.70)\\
 \midrule
 1 & \underline{7.81} (4.78) & \textbf{7.88} (4.80) & 2.57 (1.68) & 7.75 (4.92)\\
 \midrule
 2 & \underline{7.37} (4.52) & \textbf{7.43} (4.68) & 2.60 (1.77) & 6.44 (4.51)\\
 \midrule
 3 & \underline{5.84} (4.52) & \textbf{5.90} (4.62) & 2.73 (1.90) & 4.61 (5.14)\\
 \bottomrule
\end{tabular*}
\end{table}

\begin{table}[h!]
\small
\caption{\small Performance comparison of anomaly detection for the proposed optimal design, $\v d_*$, the prediction-only optimal design, $\v d_{*\text{irmse}}$, and the anomaly-only optimal design, $\v d_{*\text{sp}}$, across various scenarios.} \label{tab:spatial_sp}
\normalsize
\label{tab:st_spec} 
\centering
\begin{tabular}{llccc>{\columncolor{lightgray!30}}c}
\toprule
Design & Scenario & Specificity & Accuracy & Sensitivity & MCC \\ 
\midrule
\midrule
$\v d_{*}$ & 0 &  98.7 & 98.7 &  - &  -\\
\midrule
$\v d_{*\text{irmse}}$ & 0 & 98.9 & 98.9 &  - &  -\\
\midrule
$\v d_{*\text{sp}}$ & 0 & 99.7 & 99.7 &  - &  -\\
\midrule
\midrule
$\v d_{*}$ & 1 & \underline{99.3} & 99.0 & 67.7 & \underline{57.5}\\ 
\midrule
$\v d_{*\text{irmse}}$ & 1 & 97.9 & 97.7 & 81.8 & 47.5\\ 
\midrule
$\v d_{*\text{sp}}$ & 1 & \textbf{99.5} & 99.2 & 70.2& \textbf{63.7}\\ 
\midrule
\midrule
$\v d_{*}$ & 2 & \underline{97.1} & 96.2 & 80.4 & \underline{66.7}\\ 
\midrule
$\v d_{*\text{irmse}}$ & 2 & 95.3 & 94.6 & 83.0 & 60.7\\ 
\midrule
$\v d_{*\text{sp}}$ & 2 & \textbf{97.9} & 96.8 & 75.3 & \textbf{68.5}\\ 
\midrule
\midrule
$\v d_{*}$ & 3 & \underline{94.0} & 92.3 & 77.7 & \underline{63.6} \\
\midrule
$\v d_{*\text{irmse}}$ & 3 & 92.6 & 91.3 & 80.5 & 61.9\\
\midrule
$\v d_{*\text{sp}}$ & 3 & \textbf{95.3} & 93.4 & 76.6 &\textbf{66.8}\\
\bottomrule
\end{tabular}
\end{table}

Table \ref{tab:spatial_sp} illustrates the performance of anomaly detection among different optimal designs: $\v{d}_*$, $\v{d}_{*\text{irmse}}$, and $\v{d}_{*\text{sp}}$ across various scenarios.
Results in the table were computed with $B=15,000$ Monte Carlo draws, resulting in $B \times n = 90,000$ binary classifications. 
In scenarios where anomalies are present, we observe that the design optimised for specificity has the highest specificity, as expected.
The specificity is generally high across all designs in scenarios 0, 1 \& 2 due to a proportionally high number of True Negatives, given the class imbalance with an anomaly rate of 0\%, 1\% \& 5\%, respectively.
For a more balanced evaluation of the designs, we also consider the Matthews Correlation Coefficient (MCC), defined in Equation \eqref{eq:mcc}.
We observe that $\v{d}_*$ consistently outperforms $\v{d}_{*\text{irmse}}$, with a gain ranging between 1.4 - 10.5\% in the MCC.
The comparison study indicates a trade-off between predictive accuracy and anomaly detection performance in the dual-purpose utility, $\v{d}_*$, as expected.

\subsection{Spatio-temporal river simulation}

In this case study, we explore the spatio-temporal river network model outlined in Section \ref{sec:st_model}, simulating a river network with 300 segments with the R software package SSN \citep{ver2014ssn}.
The aim of this design scenario is to maximise prediction accuracy at $\hat{n}=56$ predetermined locations denoted by design, $\hat{\v d}$, while considering integrity of the data.
Recall that $\v y$ is the stacked vector of observations from the design locations over time, $T$, and denote $\hat{\v y}$ as the stacked vector of predictions predictions across the network over time.
The matrices $\mathbf{X}$ and $\hat{\mathbf{X}}$ represent the space-time design matrices for covariates, and $\boldsymbol{\beta}$ is the vector of regression coefficients.
Using the simple kriging approach, the posterior predictive mean is given by,
\begin{equation}
\mathbb{E}(\hat{\v y}| \hat{\v d}, \v y) = \hat{\mathbf{X}}\boldsymbol{\beta} + {\mathbf{S}}^\top \mathbf{C}^{-1}(\v y - \mathbf{X}\boldsymbol{\beta}),
\label{rq:post_mean2}
\end{equation}

where the covariance matrix $\mathbf{S}$, of dimension $n \times T$ by $\hat{n} \times T$, denotes the covariance between the observation and prediction points in space and time.
We assumed the tail-up exponential model as per Equation \ref{eq:tailup_covariance}, with the following priors:
\begin{align*}
    \sigma_0^2 &\sim \text{Uniform}(1 \times 10^{-5}, 1 \times 10^{-4}) \\
    \sigma_u^2 &\sim \text{Uniform}\left(1, \frac{3}{2}\right) \\
    \alpha_u &\sim \text{Uniform}\left(2, \frac{5}{2}\right) \\
    \phi &\sim \text{Uniform}(0, 1) \\
\end{align*}

\begin{figure}[H]
\centering
\begin{tabular}{ccc}
\subfloat[Prediction locations, $\hat{\v s}$, of design, $\hat{\v d}$, across the river network]{\includegraphics[width = 2.2in]{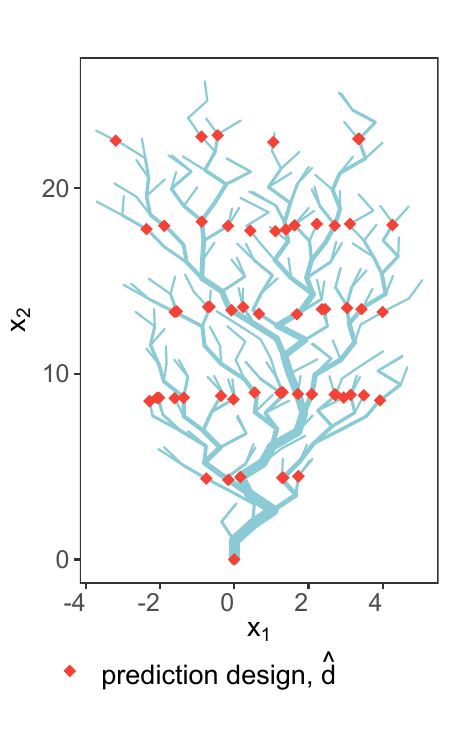}} &
\subfloat[Random start design, $\v d_1$, and the associated design space for sensor $i$ within the design.]{\includegraphics[width = 2.2in]{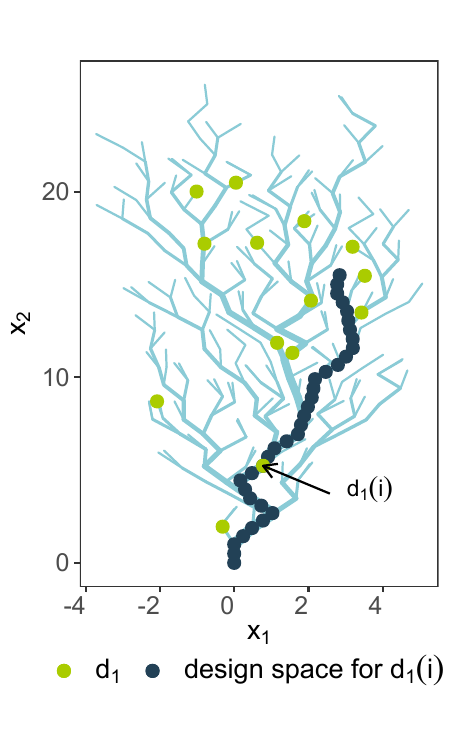}} \\
\end{tabular}
\caption{Visualisation of design configurations across a dendritic river system, comprising 300 randomly generated segments. The watercourse flows from the upper reaches toward the lower basin, converging at the river outlet positioned at coordinates (0,0).}\label{river}
\end{figure}

This design problem positions $n=14$ sensors within the river network, $\mathcal{S}$, collecting data at $T=50$ distinct time intervals.
Figure \ref{river} illustrates randomly generated design, denoted as $\v d_1$, and the prediction design, $\hat{\v d}$. 
We refer to the $i$-th coordinate of design $\v d$ as $\v d(i)$.
The intricate nature of the river network's spatial domain is evident, with the branching network topology embedded in a 3D terrestrial landscape.
To accommodate for this complexity, modifications were essential in the deployment of the ACE methodology. 
For this, we randomly assigned the design space for design coordinate, $\v d(i)$, to a corresponding network `path'. A network path is a sequence of segments spanning from the most upstream reach of a river network to the outlet. In this way, the design is governed by a singular parameter, distance upstream.
Refer to Figure \ref{river} for a visual representation of a path within the river network.

\begin{figure}[H]
\centering
\includegraphics[width = 0.6\textwidth]{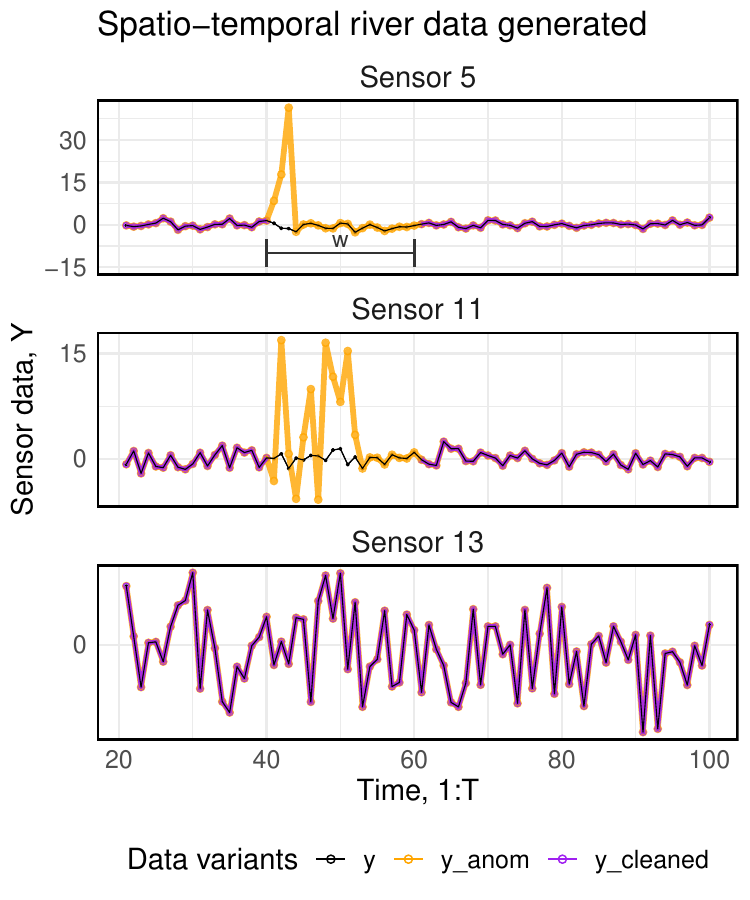}
\caption{Data variants, $\v y$, $\v y_{\text{anom}}$, and $\v y_{\text{cleaned}}$ generated in Algorithm \ref{algo:b}, for the spatio-temporal river network model, and the anomaly detection method described in Algorithm \ref{alg:Oddstream1} (see Appendix). The parameter $w$ represents the window length utilised in the anomaly detection method.}\label{fig:st_data}
\end{figure}

Consider the data variants generated in Algorithm \ref{algo:b}. 
The Oddstream anomaly detection algorithm was run with a window length of $w=50$, chosen as a hyper parameter.
Training data were independently generated from the likelihood function based on the given design.
Accordingly, each training set was comprised of $T \times n$ observations. 
For computational efficiency, anomalies were detected for all $B$ Monte Carlo draws at once.
To ensure this detection remained independent for each Monte Carlo draw, the window length, $w$, was set to $T$.
Recall that Oddstream yields an output series $(a_{1},...,a_{n})$, where $a_{i} = 1$ indicates an anomaly detected at the $i$-th sensor (otherwise, $a_{i} = 0$), for a given input window.
As such, this approach groups anomaly detection by windows of size $w$, making it effective for very large datasets and persistent anomaly patterns.
The predicted anomalies are then imputed with the posterior predictive mean, $\mathbb{E}(\hat{y}_i| \v{y}, \hat{\v d})$, for the window.
See Figure \ref{fig:st_data} for an illustration of the anomaly detection method and the resulting data variants.

\begin{figure}[H]
\begin{tabular}{ccc}
\subfloat[ACE trace plot for utility, $\text{u}$ (proposed method).]{\includegraphics[width = 0.3\textwidth]{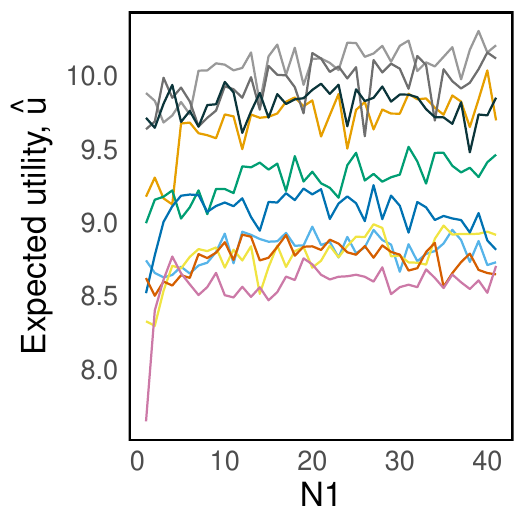}} &
\subfloat[ACE trace plot for utility, $\text{u}_{\text{irmse}}$ (prediction errors only).]{\includegraphics[width = 0.3\textwidth]{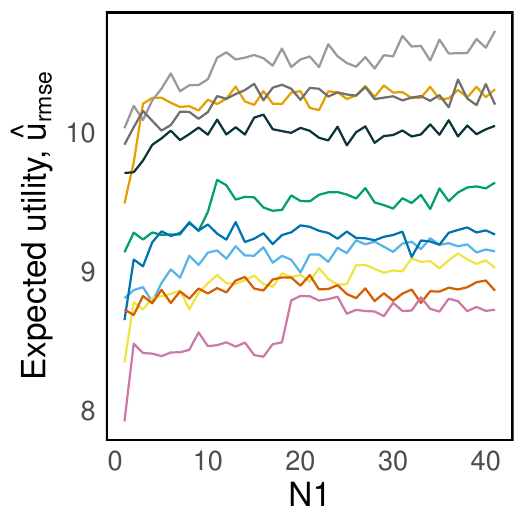}} &
\subfloat[ACE trace plot for utility, $\text{u}_{\text{sp}}$ (anomaly detection performance only).]{\includegraphics[width = 0.3\textwidth]{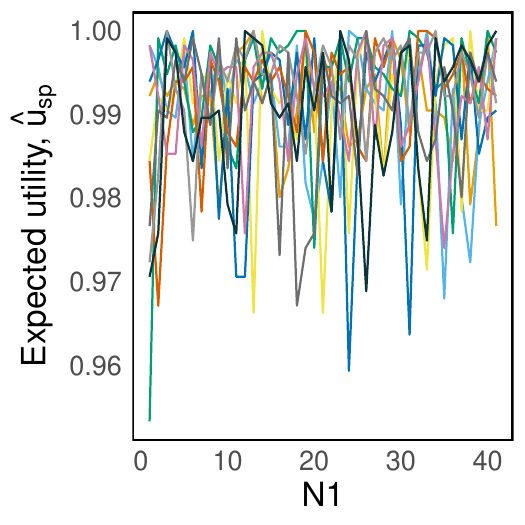}} \\
\subfloat[Locations given by optimal design, $\v d_{*}$, across the river network.]{\includegraphics[width = 0.3\textwidth]{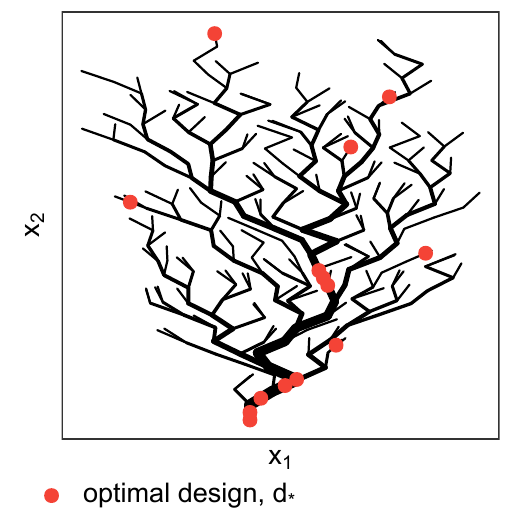}} &
\subfloat[Locations given by optimal design, $\v d_{*\text{irmse}}$, across the river network.]{\includegraphics[width = 0.3\textwidth]{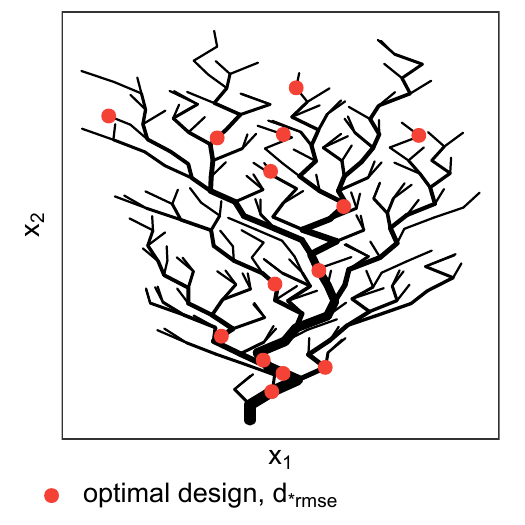}} &
\subfloat[Locations given by optimal design, $\v d_{*\text{sp}}$, across the river network.]{\includegraphics[width = 0.3\textwidth]{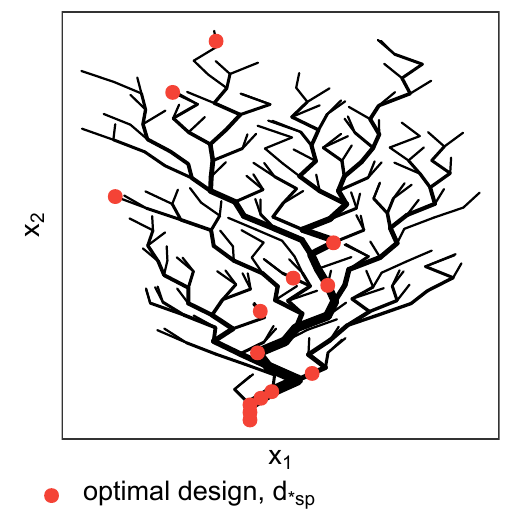}}
\end{tabular}
\caption{Top row: trace plots for the Approximate Coordinate Exchange (ACE) algorithm, optimised according to their respective utilities, defined in Equation \ref{eq:u}. Bottom row: locations of corresponding design points across the river network, for the spatio-temporal model.} \label{fig:trace_locations2}
\end{figure}

ACE was run to find optimal designs; see trace plots in Figure \ref{fig:trace_locations2}. 
At the random start, $k$, each design coordinate is assigned to a fixed `path'.
The number of random starts was set to $K=10$.
When employing an anomaly detection method that utilises windowing, the assessment of performance becomes grouped by the length of the window, reducing the number of individual data points available for evaluating performance by a factor of $w$.
Hence, we may observe increased fluctuations in the trace plot, prompting us to raise the number of iterations to $N_1 = 40$.
The locations of the optimal designs are shown in Figure \ref{fig:trace_locations2}, highlighting a similar pattern to the previous case study, even within a more complex spatial domain. 
In $\v{d}_{*\text{sp}}$, optimising for anomaly detection specificity reveals clustered optimal design locations, particularly around the river outlet.
Conversely, $\v{d}_{*\text{irmse}}$ exhibits a more uniform distribution across the river network. 
The proposed method's $\v{d}_*$ shows a mix of clustered sites around the river outlet and evenly dispersed sites across the river network.

\begin{table}[h]
\caption{\small Scenarios of various levels of anomaly generation in a spatio-temporal context, defined by the anomaly proportion, $p_{\text{A}}$, and the anomaly length, $\lambda_{\text{A}}$, for in the anomaly generation methodology in Section \ref{sec:anom_gen}. Corresponding average percentages of generated anomalies for each scenario are shown on the right.} \label{tab:scenarios} \normalsize
\begin{tabular}{@{}lllr@{}}
 \toprule
 Scenario & $p_{\text{A}}$ & $\lambda_{\text{A}}$ & \% anomalies \\ [0.5ex] 
 \midrule
 0 & 0.000 & 0.00 & 0.0\\
 \midrule
 1 & 0.001 & 1.25 & 0.6\\
 \midrule
 2 & 0.010 & 1.25 & 6.5 \\
 \midrule
 3 & 0.010 & 1.75 & 16.5 \\
 \bottomrule
\end{tabular}
\end{table}

\begin{table}[h]
\caption{\small Summary of performance metrics for the optimal designs, $\v{d}_{*}$, across different scenarios. Displayed are the expected utility, $\hat{\text{U}}(\v d_*)$, the proportion of anomalies automatically removed, and the proportion of good data retained. These results illustrate the balance between anomaly mitigation and data preservation achieved by the design across various levels of data contamination.} \label{tab:st_utility} \normalsize 
\begin{tabular*}{\textwidth}{@{\extracolsep\fill}lccc}
 \toprule
 \multirow{3}{*}{Scenario} & \multicolumn{3}{c}{$\v{d}_{*}$} \\ 
 \cmidrule{2-4}
 & \multicolumn{1}{p{2cm}}{\centering $\hat{\text{U}}(\v d_*)$} & \multicolumn{1}{p{3.5cm}}{\centering \% anomalies removed} & \multicolumn{1}{p{3.5cm}}{\centering \% good data retained} \\ 
 \midrule
 0 & 9.21 & - & 99.1 \\
 \midrule
 1 & 8.44 & 91.2 & 99.1 \\
 \midrule
 2 & 8.94 & 93.4 & 98.2 \\
 \midrule
 3 & 9.65 & 96.5 & 97.8 \\
 \bottomrule
\end{tabular*}

\end{table}

The methodology described above was executed across the scenarios detailed in Table \ref{tab:scenarios}, representing varying anomaly generation levels in a spatio-temporal context.
Table \ref{tab:st_utility} presents the proposed expected utility, $\hat{\text{U}}(\v{d}_*)$, the percentage of automatically removed (true) anomalies, and the percentage of retained good data. Notably, across all scenarios (0 to 3), both the percentage of anomalies automatically detected and the percentage of retained good data are notably high.

For instance, as the percentage of automatically removed anomalies increases from 91.2\% to 96.\%, the percentage of preserved good data remains high, over 98\% in all cases.
The findings demonstrate the ability of the optimal design to effectively identify anomalies while retaining a significant volume of good data, thereby enhancing data quality and showcasing resilience against varying contamination levels.

\begin{table}[h]
\caption{\small Efficacy of the proposed optimal design, $\v d_*$, versus the comparative designs, $\v d_{*\text{irmse}}$ and $\v d_{*\text{sp}}$, in terms of prediction error across various scenarios. Data collected at each design are assumed to be either cleaned from detected anomalies, $\v y_{\text{cleaned}}$, or anomalous, $\v y_{\text{anom}}$.} \label{tab:st_irmse}
\normalsize 
\begin{tabular*}{\textwidth}{@{\extracolsep\fill}lcccc}
 \toprule
 \multirow{2}{*}{Scenario} 
 & \multicolumn{1}{p{2.2cm}}{\centering $\text{u}_{\text{irmse}}(\v d_{*})$ w/ \\ $\mathbb{E}(\hat{\v y}| \hat{\v d}, \v y_{\text{cleaned}})$ }
 & \multicolumn{1}{p{2.7cm}}{\centering $\text{u}_{\text{irmse}}(\v d_{*\text{irmse}})$ w/ \\ $\mathbb{E}(\hat{\v y}| \hat{\v d}, \v y_{\text{cleaned}})$ }
  & \multicolumn{1}{p{2.3cm}}{\centering $\text{u}_{\text{irmse}}(\v d_{*\text{sp}})$ w/ \\  $\mathbb{E}(\hat{\v y}| \hat{\v d}, \v y_{\text{cleaned}})$ }
  & \multicolumn{1}{p{2.7cm}}{\centering $\text{u}_{\text{irmse}}(\v d_{*\text{irmse}})$ w/ \\  $\mathbb{E}(\hat{\v y}| \hat{\v d}, \v y_{\text{anom}})$ }\\ 
  \midrule
 0 & 9.43 (0.57) & \textbf{10.65} (0.63)& 9.64 (0.60) & \underline{10.63} (0.62) \\
 \midrule
 1 & 8.55 (0.53) & \textbf{8.62} (0.53) & 8.38 (0.53) & \underline{8.57} (0.53)\\
 \midrule
 2 & \textbf{9.13} (0.55) & \underline{9.10} (0.60) & 9.00 (0.56) & 8.74 (0.52)\\
 \midrule
 3 & \underline{9.79} (0.62) & \textbf{9.80} (0.63) & 9.70 (0.62) & 9.42 (0.59) \\
 \bottomrule
\end{tabular*}
\end{table}

\begin{table}[h]
\caption{\small Performance comparison of anomaly detection for the proposed optimal design, $\v d_*$, the prediction-only optimal design, $\v d_{*\text{irmse}}$, and the anomaly-only optimal design, $\v d_{*\text{sp}}$, across various scenarios.} \label{tab:st_spec} 
\normalsize
\begin{tabular}{llccc>{\columncolor{lightgray!30}}c}
 \toprule
 Design & Scenario & Specificity & Accuracy & Sensitivity & MCC \\ 
 \midrule
 \midrule
 $\v d_{*}$ & 0 & 99.1 & 99.1 & - & - \\
 \midrule
$\v d_{*\text{irmse}}$ & 0 & 96.1  & 96.1 & - & -\\
 \midrule
$\v d_{*\text{sp}}$ & 0 &  99.5  & 99.5 & - & -\\
 \midrule
 \midrule
 $\v d_{*}$ & 1  & \underline{99.1}& 98.6 & 91.2& \underline{87.4}\\
 \midrule
$\v d_{*\text{irmse}}$ & 1 & 98.5 & 98.1& 91.9 & 83.6\\
 \midrule
$\v d_{*\text{sp}}$ & 1 & \textbf{99.8}& 99.2 & 89.4& \textbf{92.0}\\
 \midrule
 \midrule
 $\v d_{*}$ & 2 & \underline{98.2} & 96.2  & 93.4 & \underline{92.2}\\ 
 \midrule
$\v d_{*\text{irmse}}$ & 2 & 96.5 &  95.5& 94.0 & 90.7\\ 
 \midrule
$\v d_{*\text{sp}}$ & 2 & \textbf{98.4}& 96.3& 93.5 & \textbf{92.4}\\ 
 \midrule
 \midrule
 $\v d_{*}$ & 3  & \underline{97.8} & 97.1& 96.5 & \underline{94.3}\\
 \midrule
$\v d_{*\text{irmse}}$ & 3 & 96.9 & 97.0& 97.0 & 93.9\\
 \midrule
$\v d_{*\text{sp}}$ & 3 & \textbf{99.8} & 97.5& 95.1 & \textbf{95.2}\\
 \bottomrule
 \bottomrule
\end{tabular}
\end{table}

Table \ref{tab:st_irmse} evaluates the prediction accuracy for the three optimal designs: $\v{d}_*$, $\v{d}_{*\text{irmse}}$, and $\v{d}_{*\text{sp}}$ across various scenarios.
Note that the sample size for the root mean squared error in this instance is up to $(\hat{n} \times T \times B)$.
Results show that $\v{d}_*$ is comparable with $\v{d}_{*\text{irmse}}$ in terms of predictive accuracy (with detected anomalies removed), while the higher errors associated with $\v{d}_{*\text{sp}}$ (detected anomalies removed) is comparable to $\v{d}_{*\text{irmse}}$ (without anomalies removed).
This demonstrates that poorly positioned sensors perform as inadequately in prediction tasks as an optimal design with contaminated data (containing anomalies of up to 16.5\%).

In Scenario 2, the proposed $\v{d}_*$ outperforms the baseline. 
However, marginal differences should not be scrutinised.
There are numerous stochastic elements influencing these outcomes, including the magnitude and duration of randomly generated anomalies, and the randomly generated training data.
Generally, we observe the near-optimal prediction accuracy of the proposed design $\v{d}_*$ compared to the baseline.

In Scenario 0, where no anomalies are introduced, $\v y_{\text{cleaned}}$ is solely influenced by false positives, with $\v y_{\text{anom}}$ being equivalent to $\v y$.
Note that, due to the anomaly indicators of Oddstream corresponding to a time window, the predicted anomalies are imputed with $\mathbb{E}(\hat{y}_i| \v{y}, \hat{\v d} )$ for the window.
As a result, we may observe lower prediction errors with $\v y_{\text{cleaned}}$ than $\v y_{\text{anom}}$ in Scenario 0.
We observe minimal errors from $\text{u}_{\text{irmse}}(\v d_{*\text{irmse}})$ w/ $\mathbb{E}(\hat{\v y}| \hat{\v d}, \v y_{\text{anom}})$ in Scenarios 0 \& 1, but as the prevalence of anomalies increases, we observe the impact that anomalies have on the prediction accuracy, as the poorest performer in Scenarios 2 \& 3.
We also observe a general increase in prediction accuracy across Scenarios 1-3, which corresponds to the general increase in sensitivity, discussed in the following.

Table \ref{tab:st_spec} reports the specificity, accuracy, sensitivity and MCC of anomaly detection given designs across different scenarios.
Again, we observe $\v{d}_*$ consistently outperforming $\v{d}_{*\text{irmse}}$, with a notable increase in MCC of up to 2.3\%, in scenario 1.
Across scenarios 1-3, $\v{d}_{*\text{sp}}$ maintains the highest specificity compared to other designs, representing the optimal design for anomaly detection specificity.
In scenario 2, MCC for $\v{d}_*$ was within 0.2 of the optimal design $\v{d}_{*\text{sp}}$.
When compared to the first case study, this case study has a more complex spatial domain, persistent anomalies across time and an advanced anomaly detection algorithm is used (which does not compute the distance between locations explicitly, but rather implicitly through the similarity of data generated at closer locations).
Nevertheless, we consistently observe the dual-utility payoff, with $\v{d}_*$ demonstrating near-optimal prediction utility and significant improvement in anomaly detection (with respect to MCC) when compared to $\v{d}_{*\text{irmse}}$.

\section{Discussion and conclusion}

Accurate and reliable data are essential for understanding natural processes, predicting environmental changes, and making informed decisions \citep{kang2016bayesian, armour2009catchment}. Policymakers, researchers, and environmental agencies often rely on river data for making critical decisions related to water management, conservation, and disaster response. 
For example, failing to distinguish between anomalies that arise from sensor malfunctions and extreme river events can lead to misdirected efforts, wasted resources, and ineffective solutions to environmental challenges \citep{leigh2019framework}.\\

In river network in-situ data, various types of anomalies can occur, indicating abnormal patterns or irregularities in the data. These anomalies can include sudden spikes or drops in water temperature, unexpected changes in flow rates, unusual levels of pollutants, or irregularities in sediment deposition. Additionally, anomalies might arise from sensor malfunctions, data transmission errors, or human interference. Detecting these anomalies is crucial for maintaining the accuracy and reliability of the data collected from river network sensors, enabling timely responses to environmental changes, pollution incidents, or equipment failures. Automated anomaly detection methods are essential for identifying these irregularities promptly and ensuring the integrity of the collected data.\\

The case study analysis highlighted the distinct strengths among the optimal design configurations: $\v{d}_*$ (proposed optimal design), $\v{d}_{*\text{irmse}}$ (prediction-focused optimal design), and $\v{d}_{*\text{sp}}$ (optimal design accounting for anomalies). 
In both case studies the optimal design, $\v{d}_*$, demonstrated high predictive accuracy similar to $\v{d}_{*\text{irmse}}$.
Conversely, the comparable optimal design, $\v{d}_{*\text{sp}}$, demonstrated high anomaly detection performance but revealed limited predictive ability, especially with increased anomalies.
The proposed design $\v{d}_*$ significantly improved anomaly detection performance compared to the baseline design $\v{d}_{*\text{irmse}}$.
Overall, we found that $\v{d}_{*}$ offers a balanced performance in both prediction and anomaly detection.\\

In the context of anomaly detection for data quality assurance, high specificity means that the anomaly detection algorithm is highly effective at correctly identifying non-anomalous data points as normal.
This is important because it minimises the risk of mistakenly discarding or altering normal data, amidst the complexity of potential anomalies in data collected from in-situ sensors, which may have diverse origins.
However, overemphasising specificity can lead to a model that misses anomalies in the data, potentially missing critical anomalies in the process. 
In cases like extreme weather events, where anomaly certainty is crucial, metrics like accuracy may be more relevant. In our context, the focus on specificity helped to filter out some outliers, but it's essential to consider other anomaly detection metrics depending on the scenario. Future work may also extend the approach to allow for the potential of missing data within the design.\\

While our focus has been on BOED for prediction and anomaly detection for river network sensor data, there are broader implications and applications of this novel method. Spatial and spatio-temporal models are extensively applied across a wide range of research fields, from economics to environmental science, urban planning, epidemiology, and meteorology. 
Optimal design is crucial in collecting and analysing data, providing insights into complex dynamics and interactions in space and time. 
Intelligent design with anomaly detection is an exciting new concept that can provide invaluable tools for researchers and policymakers more broadly, from tracking the spread of diseases, planning more effective urban development strategies, to better predicting extreme weather patterns.

\section{Acknowledgments}
This project was supported by the Australian Research Council (ARC) Linkage Project (LP180101151) ``Revolutionising water-quality monitoring in the information age". 

\section{Data and Code Accessibility}
The design optimisations were performed using the \texttt{R} package \texttt{acebayes} \citep{overstall2017acebayes}.
This code implementation can be found at \url{https://github.com/KatieBuc/design_anom}.

\begin{appendices}

\section{Extended Data}\label{secA1}

\subsection{Oddstream Anomaly Detection}\label{oddstream_app}

\begin{algorithm}[H]
\tcp{Training Phase}
\caption{Oddstream: Anomaly Detection in Spatio-temporal Data}\label{alg:Oddstream1}
\KwIn{Collection of time series data, $\v y_{\text{train}} \in \mathbb{R}^{l_{\text{train}} \times n}$, representative of the typical system behavior.}
\KwOut{Anomalous threshold, $\tau^*$.}
\BlankLine
Extract $m$ features from each time series in $\v y_{\text{train}}$ and create a feature matrix $\mathbf{M}$ of size $n \times m$.\;
Normalise the columns of $\mathbf{M}$ to produce $\mathbf{M}^*$.\;
Apply principal component analysis to $\mathbf{M}^*$.\;
Create a 2D space using the first two principal components; each data point in this space corresponds to a time series in $\v y_{\text{train}}$.\;
Estimate the probability density, $\hat{f}_2$, of the 2D space using kernel density estimation with a bivariate Gaussian kernel.\;
Draw a large number of extremes from the estimated probability density function, $\hat{f}_2$, and form an empirical distribution in the $\Psi$-transform space, with
\[\Psi|\hat{f}_2(x)| = 
\begin{cases}
(-2ln(\hat{f}_2(x))- 2ln(2\pi))^{\frac{1}{2}}, & \hat{f}_2(x) < (2 \pi)^{-1} \\
0, & \text{otherwise}.
\end{cases}
\] \;
Fit a Gumbel distribution to the $\Psi$-transformed values.\;
Determine the anomalous threshold contour, $\tau^*$, (i.e. where the most extreme of the $n$ typical samples generated from $f_2$ will lie to some given level of probability, $\tau_F$) by equating the CDF of the fitted Gumbel distribution to $\tau_F$.\;

\BlankLine
\tcp{Online Anomaly Detection Phase}
\KwIn{Matrix of data as a sliding window $\v y_{\text{anom}, [j - w, j]} \in \mathbb{R}^{w \times n}$, anomalous threshold $\tau^*$.}
\KwOut{Indicator vector, $\v{a}_j$, of anomalous series in $\v y_{\text{anom}, [j - w, j]}$.}
\BlankLine
Extract $m$ features from the time series $\v y_{\text{anom}, [j - w, j]}$ to create feature matrix $\mathbf{M}_{\text{test}}$.\;
Project $\mathbf{M}_{\text{test}}$ onto the 2D PC space obtained from typical data.\;
Calculate probability densities for the projected data points with respect to $\hat{f}_2$ (from the training phase).\;
Identify time series where $\hat{f}_2(y_i) < \tau^*$ for $i = 1, 2, \ldots, n$ and return $\v{a}_j = (a_{j1},...,a_{jn})$ with $a_{ji} = 1$ for an anomaly in the $i$-th sensor (else $a_{ji} = 0$), given window input $\v y_{\text{anom}, [j - w, j]}$.\;
Repeat steps 1–4 for each new time window.\;
\end{algorithm}

\subsection{Approximate coordinate exchange}\label{ace_app}

We specify a design, $\v d \in \mathcal{D} \subset \mathbb{R}^{n \times k}$.
Phase I of the algorithm utilises cyclic ascent to maximise the approximation $\hat{\text{U}}(\v d)$ to the expected utility. For each coordinate, $x_{ij}$, a one-dimensional emulator of $\tilde{\text{U}}(\v d)$ is constructed as the mean of the posterior predictive distribution conditioned on a few evaluations of $\hat{\text{U}}(\v d)$, assuming a Gaussian process (GP) prior.
The emulator for the $ij$-th coordinate is constructed as follows:
\begin{enumerate}
    \item Select a one-dimensional space-filling design with points, $x^q_{ij} \in \mathcal{D}_{ij}$, for $q = 1, \ldots, Q$.
    \item Construct designs, $\v d^q_{ij}$, identical to the current design but with the $ij$-th coordinate replaced by a point in the one-dimensional space-filling design,
    $$\v x^q_{i} = (x_{i1}, \ldots, x_{ij-1}, x^q_{ij}, x_{ij+1}, \ldots, x_{ik})^\top.$$
    \item Evaluate $\tilde{\text{U}}(\v d^q_{ij})$ for $q = 1, \ldots, Q$.
    \item Fit a GP regression model to the data $\{\v x^q_{i}, \tilde{\text{U}}(\v d^q_{ij})\}_{q=1}^{Q}$ and set emulator $\hat{\text{U}}_{ij}(x)$ as the resulting predictive mean.
    \item Find $$x^*_{ij} = \text{arg max}_{x \in \mathcal{D}_{ij}}\hat{\text{U}}_{ij}(x)$$
    and produce design $\v d^*_{ij}$ with $i$-th row $(\v x^*_{i})^\top = (x_{i1}, \ldots, x_{ij-1}, x^*_{ij}, x_{ij+1}, \ldots, x_{ik})^\top$.
\end{enumerate}
The approach to maximising $\hat{\text{U}}_{ij}(x)$ is robust to multi-modal emulators and is efficient due to the minimal computational overhead in evaluating the predictive mean.

\end{appendices}

\bibliography{sn-bibliography}

\end{document}